\documentclass[journal=jacsat,manuscript=article]{achemso}
\usepackage{graphicx}
\usepackage{dcolumn}
\usepackage{bm}
\usepackage{xspace}
\usepackage{amsmath,amssymb}
\usepackage[utf8]{inputenc}
\usepackage[T1]{fontenc}
\usepackage{mathptmx}
\usepackage{etoolbox}
\usepackage{xcolor}

\makeatletter
\def\@email#1#2{%
 \endgroup
 \patchcmd{\titleblock@produce}
  {\frontmatter@RRAPformat}
  {\frontmatter@RRAPformat{\produce@RRAP{*#1\href{mailto:#2}{#2}}}\frontmatter@RRAPformat}
  {}{}
}%
\makeatother

\def\mos{${\rm MoS_2}$}
\def\Wmk{\ensuremath{\rm\, W/mK}\xspace}

\author{Riccardo Dettori}
\affiliation{Department of Physics, University of Cagliari, Monserrato, CA, 09042 Italy}
\email{riccardo.dettori@dsf.unica.it}
\author{Francesco Siddi}
\affiliation{Department of Physics, University of Cagliari, Monserrato, CA, 09042 Italy}
\author{Luciano Colombo}
\affiliation{Department of Physics, University of Cagliari, Monserrato, CA, 09042 Italy}
\author{Claudio Melis}
\affiliation{Department of Physics, University of Cagliari, Monserrato, CA, 09042 Italy}

\title{Enhancing heat transport in MoS$_2$ via defect-engineering.}

\begin{document}

\begin{abstract}
{\mos \ is one of the most investigated and promising transition-metal dichalcogenides. Its popularity stems from the interesting properties of the monolayer phase, which can serve as the fundamental block for numerous applications. In this paper, we propose an atomistic perspective on the modulation of thermal transport properties in monolayer \mos \ through strategic defect engineering, specifically the introduction of sulfur vacancies. Using a combination of molecular dynamics simulations and lattice dynamics calculations, we show how various distributions of sulfur vacancies--ranging from random to periodically arranged configurations--affect its thermal conductivity. Notably, we observe that certain periodic arrangements restore the thermal conductivity of the pristine system, due to a minimized interaction between acoustic and optical phonons facilitated by the imposed superperiodicity. This research deepens the understanding of phononic heat transport in two-dimensional materials and introduces a different point-of-view for phonon engineering in nanoscale devices, offering a pathway to enhance device performance and longevity through tailored thermal management strategies.}
\end{abstract}

\maketitle

Transition metal dichalcogenides (TMDCs) are characterized by strong in-plane bonding and weak out-of-plane interactions, enabling their exfoliation into two-dimensional (2D) layers with single unit cell thickness\cite{Novoselov2005}. Represented by the formula MX$_2$, where 'M' denotes a transition metal from group IV, V, or VI, and 'X' represents a chalcogen (S, Se, or Te), these compounds consist of layered structures with two hexagonal planes of chalcogen atoms and a plane of metal atoms in between. Recent advancements in nanoscale materials characterization and device fabrication have opened new opportunities for applying thin TMDC layers in nanoelectronics and optoelectronics \cite{LopezSanchez2013,Splendiani2010}. \mos \ has been extensively investigated for its potential in field-effect transistors, making it a promising candidate for low-power, high-performance electronic devices\cite{Radisavljevic2011,Zhang2012}. Current trends focus on combining \mos \ with other 2D materials to form heterostructures aimed at advancing technologies in quantum computing and sensing\cite{Manzeli2017}. 

{Effective thermal management is a key element for the performance and longevity of \mos-based electronic and optoelectronic devices. For example, the creation of heterostructures with other 2D materials, such as graphene, can enhance thermal transport properties, with the interfaces serving as efficient thermal pathways\cite{Ghosh2010}. } Following the same trend of other physical properties (bandgap\cite{Mak2010}, Young modulus\cite{Samy2021}, etc.), the thermal conductivity of \mos \ is significantly affected when transitioning from bulk to monolayer forms. While the bulk phase has a relatively low thermal conductivity due to its layered structure and weak interlayer van der Waals interactions, which hinder efficient heat transfer, monolayer \mos, shows higher in-plane thermal conductivity due to the absence of these interlayer forces and the stronger covalent bonding within the layer. {Furthermore, the layered structure of the bulk phase results in low-frequency optical phonon branches near the acoustic ones, lowering the overall conductivity\cite{Gandi2016}, and in addition yield a strongly anisotropic thermal conductivity tensor.} Experimental reports indicate in-plane thermal conductivities for monolayer \mos \ ranging from $13\Wmk$ to $54.5\Wmk$ depending on the measurement conditions and the sample synthesis methods\cite{Yan2014,Bae2017,Gandi2016}. {Similarly, contradictory theoretical results ranging from $1.35\Wmk$ to $140\Wmk$ highlight the significant influence of computational methodologies on the calculation of thermal conductivity\cite{Varshney2014,Liu2013,Jiang2013,Li2013,Cai2014}.}

The moderate thermal conductivity and high electrical conductivity of \mos \ suggest potential applications in the field of thermoelectric conversion. Strategies to optimize the thermoelectric figure of merit $\rm ZT$ of \mos \ include doping and strain engineering\cite{Huang2013, Gabourie2022}. In this context, nanostructuring monolayers along with tailored defect engineering, can significantly alter lattice thermal conductivity\cite{Dettori2015,Dettori2015b,Lorenzi2018,Mahendran2024}. Defects, such as point defects (vacancies, interstitials, antisites), line defects (grain boundaries), and surface defects, break the lattice symmetry and serve as scattering centers for phonons. Sulfur vacancies are more common and can significantly reduce thermal conductivity\cite{Zhou2013}. Extended defects can also act as phonon barriers, effectively impacting thermal conductivity\cite{Lorenzi2018,Sledzinska2016,Muratore2014}. However, despite extensive research \cite{Ding2015,Yarali2017,Chen2020,Wang2016,Polanco2020}, there is still limited knowledge on how defect distribution affects thermal transport properties of this material. Understanding and tailoring defect distributions can effectively tune thermal conductivity and enhance heat management.

In this work, we adopt a combination of molecular dynamics (MD) simulations and lattice dynamics (LD) calculations to investigate how varying distributions of sulfur vacancies can influence the thermal transport properties of \mos. In particular, we first analyze the effects of randomly arranged vacancies at different concentrations. Next, we reorganize the same concentration of defects into regular patterns or, in other words, realizing ordered strips of vacancies, indeed creating superlattice-like (SL) structures. {Our findings reveal that the thermal conductivity is influenced by the spacing of these strips (the period of the resulting SL), and for certain values of the SL period, the thermal conductivity matches that of the pristine material. In particular, phonon analysis reveals that for a specific spacing of the S-vacancy strips, the acoustic phonons in \mos \ remain largely unaffected, with reduced interaction with optical phonons. Consequently, phonon group velocities are only slightly affected, leading to a negligible reduction in thermal conductivity.}

To investigate thermal transport, we adopt the Approach-to-Equilibrium Molecular Dynamics (AEMD) in our MD simulations. AEMD is a robust technique for calculating thermal conductivity, primarily due to its ability to reduce the long simulation times typically required by other non-equilibrium and equilibrium methods. While detailed technical descriptions are available elsewhere\cite{Melis2014,Cappai2024}, we will briefly outline the core methodology here. Assuming the absence of heat sources or sinks and no convective transport within the {simulated specimen of length $L_x$}, we impose periodic boundary conditions and a step-like temperature profile {along the $x$ direction}
\begin{equation}\label{HEAT4-MD}
 T(x,t=0)=\begin{cases}
   ~T_{\text H} ; \quad 0 \le x\le \frac{L_x}{2}\\
   ~T_{\text C} ; \quad \frac{L_x}{2} < x < {L_x} 
\end{cases} 
\end{equation}
In this context, the one-dimensional heat equation solution can be expressed as the average temperature difference between the two regions of the sample
\begin{equation}\label{HEAT6B-MD}
 \langle \Delta T(t) \rangle=\langle T_\text H(t)\rangle-\langle T_\text C(t)\rangle=\frac{8\langle \Delta T(0) \rangle}{\pi^2}\sum_{n=0}^{\infty} \frac{1}{2n+1}e^{-\beta_n^2\bar{\kappa} t}
 \end{equation}
where $T$ is system temperature, $\beta_n=\pi n/L_x$, and$\bar{\kappa}$ is the thermal diffusivity. Implementing AEMD in MD simulations thus requires the realization of a step-like initial state, through two separate NVT runs for the two regions of the simulation cell, followed by its evolution in the microcanonical ensemble. The kinetic temperature computed during the simulation is fitted with Eq.~\eqref{HEAT6B-MD}, using the thermal diffusivity $\bar{\kappa}$ as the fitting parameter, to eventually yield thermal conductivity via $\kappa=\bar{\kappa} \rho \mathcal{C}_v$, where $\rho$ is the system mass density and $\mathcal{C}_v$ is its specific heat.

Phonon properties, including eigenmodes, group velocities, and dispersion relations, were obtained using the harmonic approximation within the frozen phonon method as implemented in the Alamode suite\cite{Tadano2014}. Both MD simulations and LD calculations have been performed using the LAMMPS package\cite{Thompson2022}, employing the modified Stillinger-Weber potential for TMDCs and \mos\cite{Jiang2015,Jiang2018}. This interaction scheme, while not specifically designed for thermal properties, provides good agreement with ab initio phonon spectra and ensures stable simulations for defected \mos \ structures. Equations of motion were integrated using the velocity Verlet algorithm with a timestep of 0.5 fs.
\begin{figure}[htp]
\begin{center}
\includegraphics[width=0.85\textwidth]{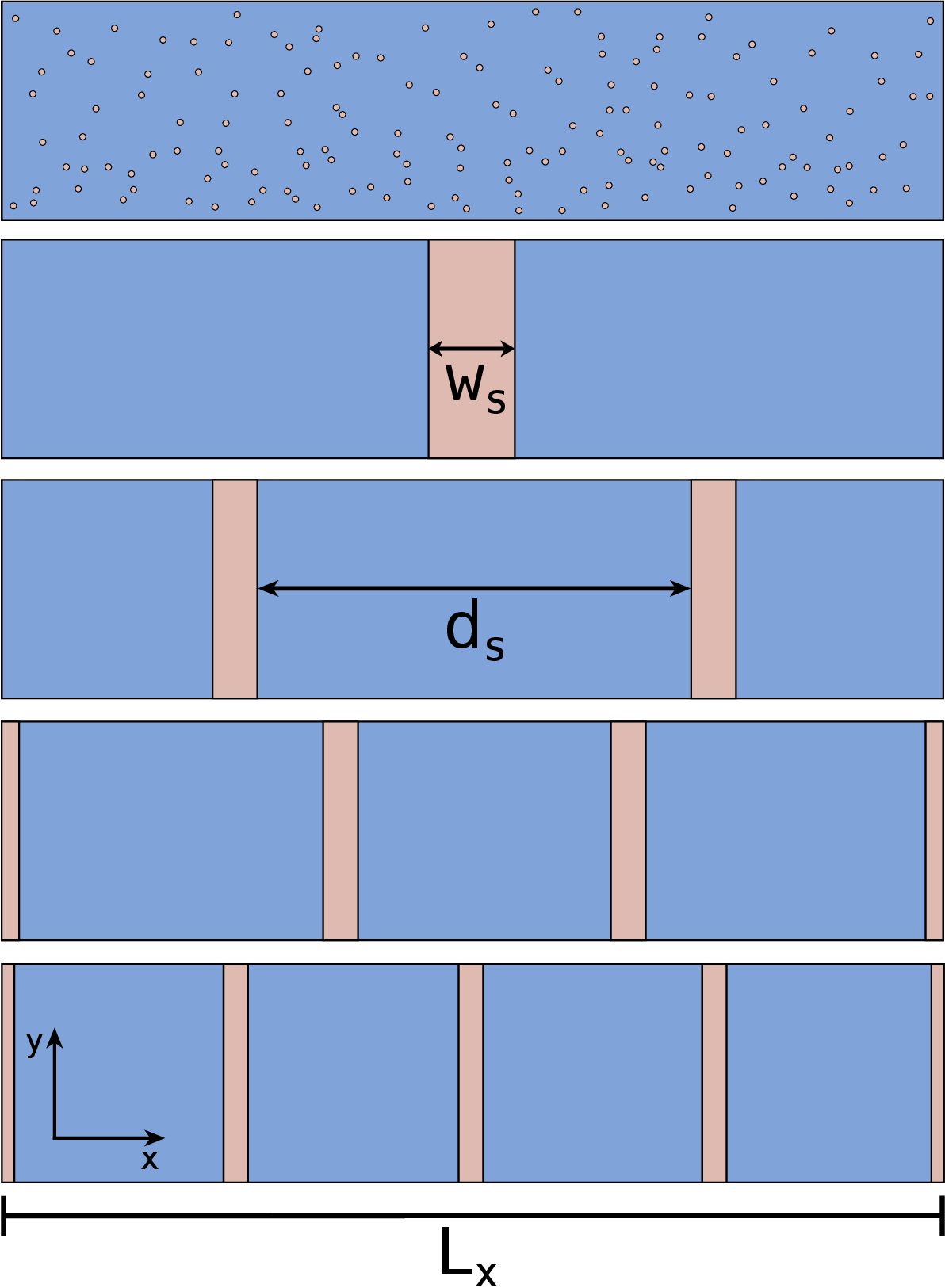}
\caption{Schematization of some of the system here investigated: pink spots/regions identify the missing sulfur atoms, while blue areas represent the pristine regions.}
\label{fig:defectedmos2}
\end{center}
\end{figure}

Starting from a rectangular conventional cell with optimized lattice constants $a=3.1175\rm$ \AA{} and $b=5.3997$ \AA{}, we introduced random sulfur vacancies by removing single sulfur atoms only from the top layer, {in order to simulate etching processes where liquid chemicals, reactive gases or plasma are used to selectively remove portions of material from the surface\cite{Huang2013}}. The vacancy densities investigated were $\rho_S=1\%,2.5\%,5\%,$ and $10\%$  of the total number of top sulfur atoms. A conceptual representation of the resulting structures is reported in the top panel of Fig.~\ref{fig:defectedmos2}: the pink regions represent missing sulfur atoms. For the ordered distributions, we created strips of sulfur atoms corresponding to the same densities as the random samples. The width of each strip, $w_s$, is determined by:
\begin{equation}
    w_s=\dfrac{h}{n_s}\dfrac{L_x}{a}\rho_s
\end{equation}
where $h=3.1387$ \AA{} is the minimum distance between two adjacent sulfur atoms in the $x$ direction, $n_s$ is the number of strips realized in the sample, $a$ is the lattice constant, $L_x$ is the system length in $x$ direction in crystal units and $\rho_s$ is the sulfur vacancies density. It follows that: (a) to realize the same number of vacancies, a larger number of strips results in a smaller $w_s$, and (b) the ratio $h/a\approx 1$, indicating that $\rho_s$ provides direct information on the relation between $w_s$ and $L_x$. Finally, the distance between strips is given by $d_s=(L_x-w_s)/n_s$.

For our AEMD simulations, we realized \mos \ monolayer by replicating the aforementioned conventional cell $L_x \times 16b$ times, with $L_x$ ranging from $100a$ to $2000a$, thus spanning the range $31.2$--$623.5$ nm. To avoid spurious results due to interactions with periodic images, we set $L_z=5$ nm. Configurational averages were obtained by conducting multiple statistically independent simulations, varying from $20$ replicas for the smaller samples to $2$ replicas for the longest one, each with different initial atomic velocities. Furthermore, since thermal conductivity is highly dependent on the distribution of defects, the random systems were initialized with different vacancy sites for each replica. The resulting simulation cell requires careful preparation, to ensure stability during production runs. Initially, atomic geometries were optimized by allowing the in-plane lattice constants to relax. A progressive annealing from $100$ K up to $300$ K was then performed in order to guarantee a complete equilibration of the sample. The step-like configuration was established through two separate 10 ps-long NVT runs (with $T_H=325$ K and $T_C=275$ K), while the NVE duration varied with the system length, ranging from 350 ps up to 7 ns.

\begin{figure}[htp]
\begin{center}
\includegraphics[width=0.85\textwidth]{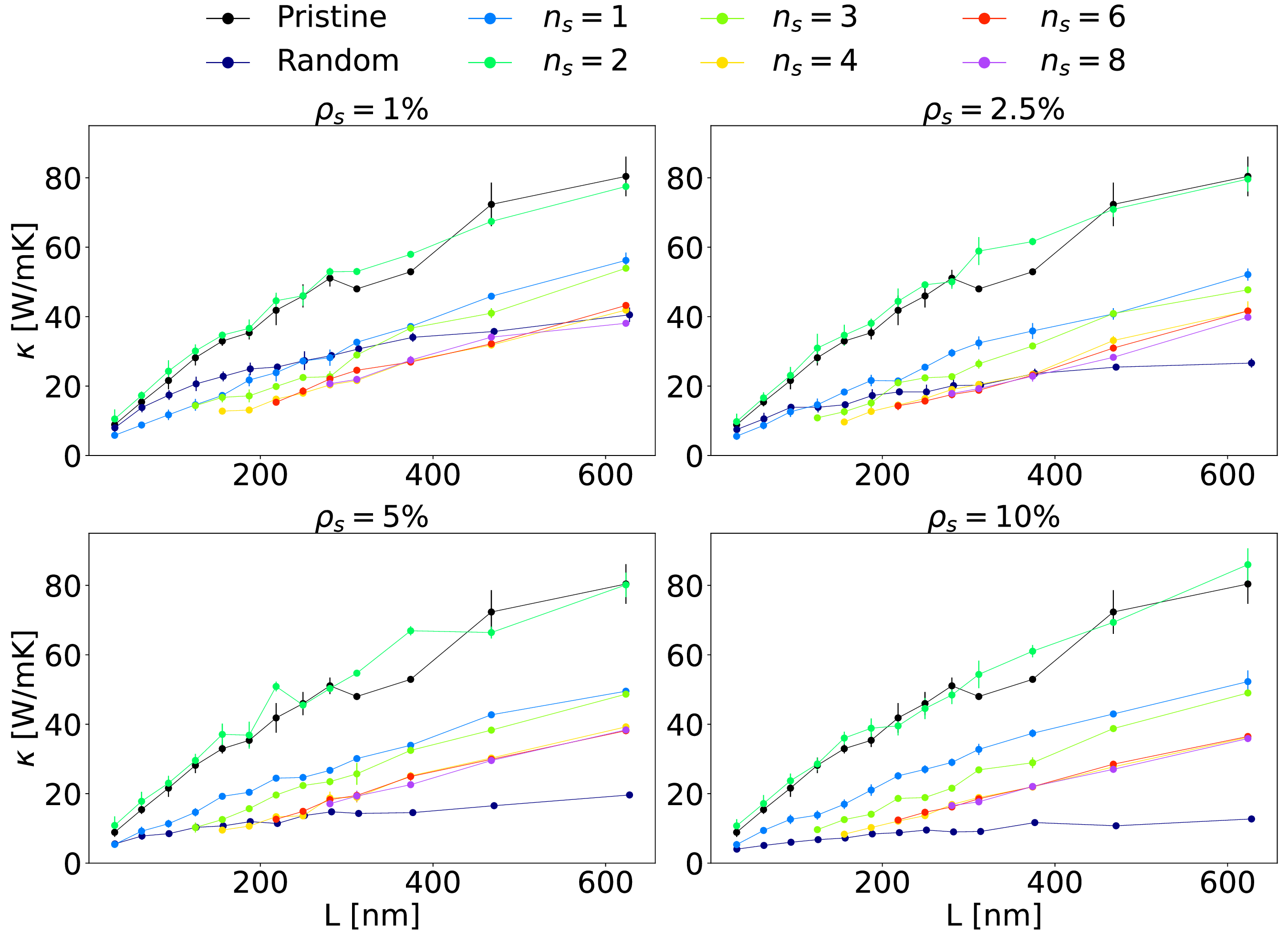}
\caption{Thermal conductivity of defected \mos \ as a function of system length, for different numbers of sulfur vacancies ($\rho_S$) and for different distributions, from random defects to $n_s=1,2,3,4,6,$ and $8$ ordered strips. The error bars are calculated as the standard error from configurational averages.}
\label{fig:k_vs_L}
\end{center}
\end{figure}

Figure~\ref{fig:k_vs_L} shows the AEMD results as a function of the different parameters here considered: $L_x$, $n_s$, and $\rho_s$. {Although we are well aware of size effects in non-equilibrium methods (which in fact generates the extensive range of theoretical/computational thermal conductivity values reported \ in literature), we refrained from extrapolating $\kappa_\infty$ for both pristine and defected systems, as it lies beyond the scope of this study. Our findings indicate a similar length dependence in ordered samples as observed in the pristine system. For this reason, in what follows we choose to reference the conductivity values of defected specimens to their pristine counterparts for a specific system size.} As anticipated, samples with random defects exhibit reduced $\kappa$ which tends to saturate when longer systems are considered. The introduction of random sulfur vacancies increases phonon scattering and limits their mean free path (MFP), as evident from the stronger reduction in $\kappa$ with increasing $\rho_s$. On the other hand, for ordered strips of vacancies, $\kappa$ is reduced but with two key differences: {counterintuitively, thermal transport up to the lengths here considered exhibit an appreciable size effect}, and increasing the number of sulfur vacancies (i.e. increasing $w_s$) does not further decrease thermal conductivity. This suggests that the SL periodicity introduced by ordered vacancy regions influences thermal transport in a non-trivial manner.
\begin{figure}[htp]
\begin{center}
\includegraphics[width=0.85\textwidth]{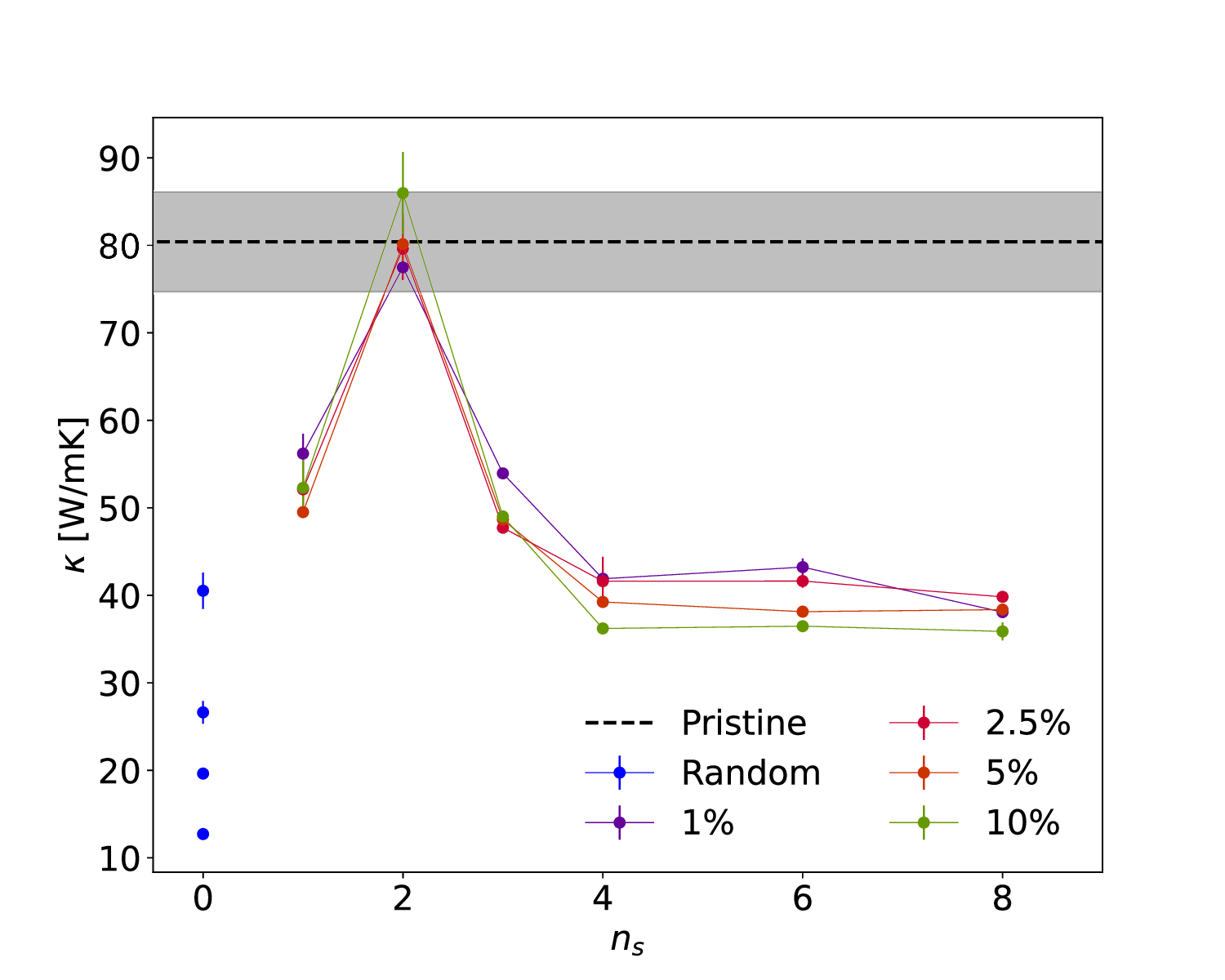}
\caption{Thermal conductivity of different defected \mos \ distributions and defect concentrations for a $L_x=623.5$ nm-sample. The shaded area represents the standard error associated with the computed $\kappa$ of the pristine monolayer.}
\label{fig:k_vs_periodicity}
\end{center}
\end{figure}
To further explore this feature, Fig.~\ref{fig:k_vs_periodicity} summarizes our finding for $L_x=623.5$ nm as a function of \mos \ monolayer super periodicity $n_s$. Except for $n_s=2$, $\kappa$ weakly depends on the super periodicity and $\rho_s$, contrary to random samples. However, not only a systematic maximum in thermal conductivity is observed for $n_s=2$, but the value of $\kappa$ of pristine \mos \ is recovered within the uncertainty regardless of the number of vacancies. We remark that this effect is observed for any $\rho_s$ and at any length here investigated. 
\begin{figure}[htp]
\begin{center}
\includegraphics[width=0.85\textwidth]{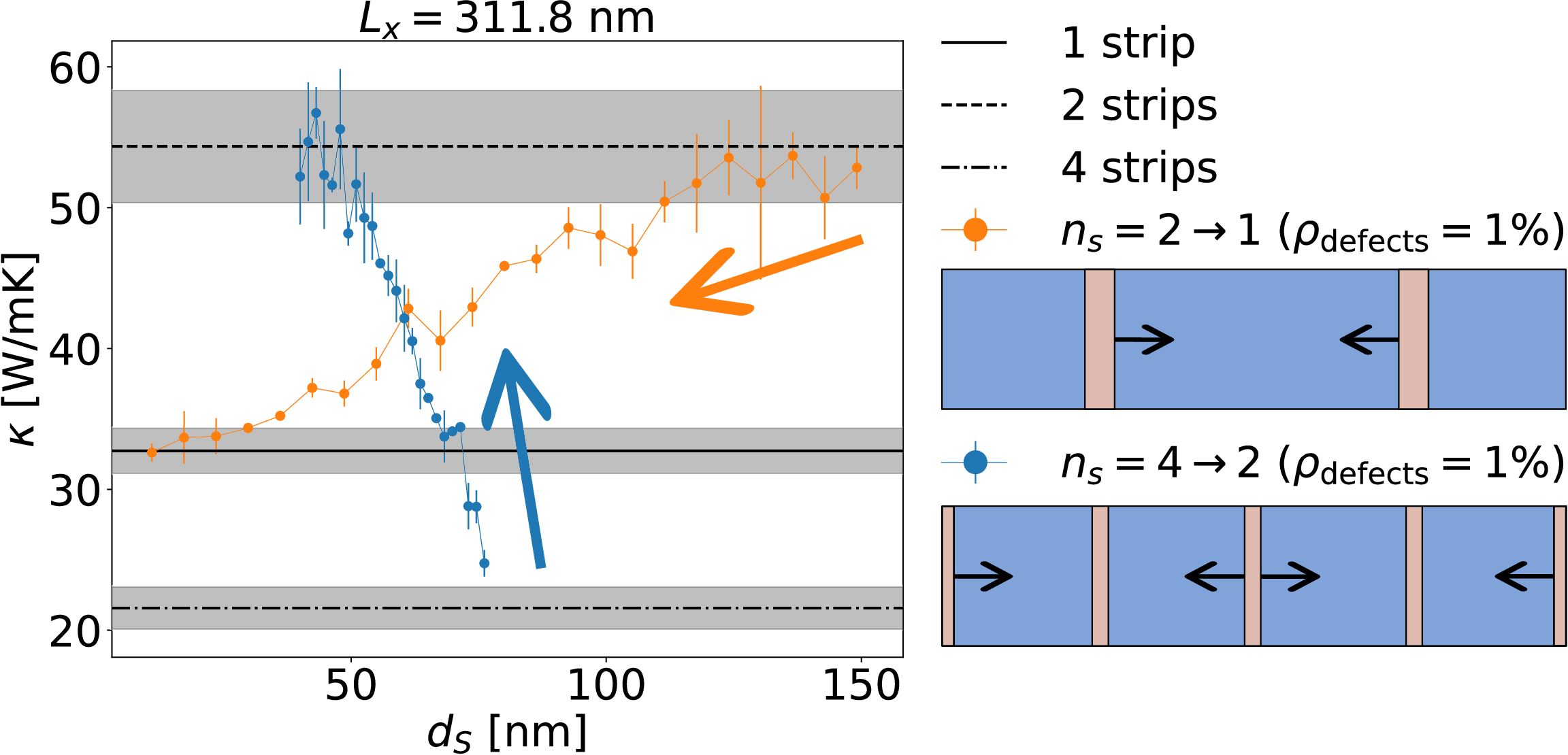}
\caption{The effect of merging two strips of vacancies into a single one or four strips into two equispaced ones in  a $L_x=311.8$ nm-sample, according to arrows in the plot and the scheme reported on the right. The shaded areas represent the standard errors associated with the computed $\kappa$ for those specific configurations.}
\label{fig:k_merging}
\end{center}
\end{figure}

{Generally, the reduction in thermal conductivity correlates directly with a decrease in phonon MFP due to the introduction of scattering centers such as defects and interfaces. However, our results indicate that in ordered samples, superperiodicity could play a major role. Specifically, for a given SL period ($n_s = 2$), phonons with certain heat-carrying wavelengths can propagate unimpeded through the system.}
This guess is confirmed by varying the relative distance between strips and observing the changes in thermal conductivity (Figure~\ref{fig:k_merging}). The thermal conductivity depends on the relative strip distance monotonically, and the $\kappa$ value for $n_s = 1$ is achieved when two strips are separated by $d_s \lesssim 20$ nm. Similarly, when $4$ strips are brought closer to form $2$, the conductivity for $n_s = 2$ is reached already when adjacent strips are close but not coalescent, like the perfectly equispaced case. In other words, a minimum distance exists where phonons propagating along the $x$ direction perceive separate columns as a single one, efficiently transmitting into the next pristine region. While it is widely known that realizing SLs combining together two or more different materials represents an efficient strategy for phonon engineering and creating phonon barriers\cite{Wang2020,Dettori2015b,Sood2021}, the realization of structures that enhance phonon propagation to the point that the thermal conductivity of the SL matches the one of the host/pristine material is not straightforward. It has been shown that interface quality and SL periodicity can increase thermal conductivity due to coherent phonon transport and constructive phonon scattering at interfaces \cite{Guo2018,Malhotra2019,Luckyanova2012}. Since also the arrangement of phonon bands in the constituent materials can affect how phonons are transmitted or reflected at interfaces, we performed LD calculations on prototypical SL structures. Although our LD calculations are performed adopting the same interatomic potential as the MD simulations, the introduction of defects disrupts the original crystalline symmetry, significantly reducing the number of symmetry operations allowed to build the dynamical matrix and resulting in a large number of atomic displacements and interatomic force constants. This significantly limits the system sizes for these calculations. Thus, we used a simulation cell with a planar size of $8a\times 2b$ ($L_x=2.5$ nm and $L_y=1.1$ nm) containing a total of 96 atoms from which we removed 6 sulfur atoms from the upper layer realizing a random distribution and ordered distributions using $n_s=1,2,$ and $3$ strips.

\begin{figure*}[htp]
\begin{center}
\includegraphics[width=0.85\textwidth]{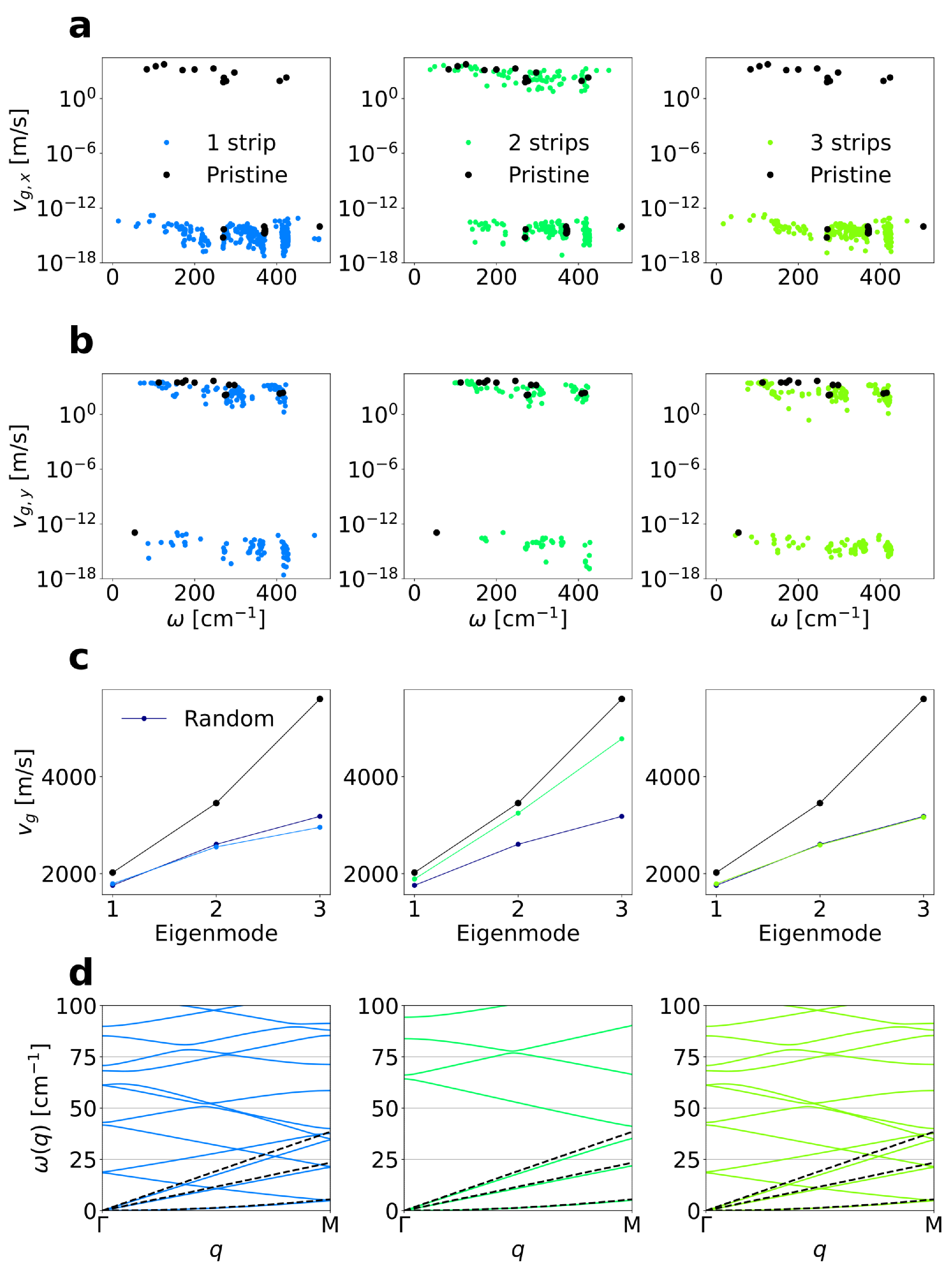}
\caption{(a,b) Mode-averaged phonon group velocities along $x$ and $y$ directions, respectively. (c) Frequency-averaged group velocity for the three acoustic branches of pristine and defected \mos. (d) Phonon dispersion curves along the $\Gamma\rightarrow M$ direction. Black dashed lines represent the acoustic branches of the pristine monolayer.}
\label{fig:lattice_dynamics}
\end{center}
\end{figure*}

The results are reported in Fig.~\ref{fig:lattice_dynamics}. Panel a and b show mode-averaged phonon group velocities ($v_{x,y}(\omega)=\frac{1}{m}\sum^m_{\mathbf{q}j} \mathbf{v}_{\mathbf{q}j}(\omega)$) along $x$ and $y$ directions, respectively. The data, compared with pristine \mos \ show that defects significantly reduce $v_x$ in all configurations except for $n_s=2$, where a considerable number of modes maintain non-zero group velocities. {On the other hand, the plots show basically no differences in $v_y$, suggesting that thermal transport along the direction parallel to the strips is unaffected.} In panel c, we focus on the frequency-averaged group velocity $\langle v_g \rangle$ of the acoustic modes. Acoustic phonon modes are generally the most responsible for thermal transport in non-metallic solids, due to their lower frequency and longer wavelength compared to optical phonons\cite{Callaway1959,Broido2007,Pop2010}. Consistent with observations in Figs.~\ref{fig:k_vs_L},~\ref{fig:k_vs_periodicity}, the group velocity of acoustic modes in defected structures is greatly hindered compared to the pristine system, roughly halving the $\langle v_g \rangle$ {and exhibiting values comparable with the random distribution samples}. Again, in $n_s=2$ systems, it appears that these modes are only marginally affected with $\langle v_g \rangle$ comparable with their pristine counterparts. This phenomenon is better visualized by considering phonon dispersion relations $\omega(q)$ following heat transport direction $\Gamma\rightarrow M$, as shown in Fig.~\ref{fig:lattice_dynamics}d: the introduction of defects is reflected in a large number of localized vibrational modes in the form of low-frequency optical branches, which can scatter with acoustic modes thus reducing their MFP and lifetimes. As a matter of fact, $n_s=1$ and $3$ samples display a large number of bands crossing spots, {resulting in a decrease of acoustic phonon group velocities and the creation of additional channels for phonon-phonon scattering}\cite{Siddi2024,Li2016,Christensen2008}. Acoustic phonons can scatter into optical phonons and vice versa, increasing the overall scattering rate\cite{Chen1997}. When the vacancies are arranged with periodicity $n_s=2$ instead, localized defect modes occur at higher frequencies, essentially realizing a phonon band gap that reduces the scattering events between acoustic and optical phonons, since fewer states are available for scattering thus enhancing their MFP. Furthermore, Umklapp processes involving interactions between phonons of different branches are strongly suppressed, greatly reducing the system thermal resistance\cite{Maznev2014}. As displayed pretty clearly in the 2nd plot of Fig.~\ref{fig:lattice_dynamics}d, acoustic phonons are essentially unchanged with respect to their pristine \mos \ equivalent. This is in agreement with a similar result in short-period SiGe SL\cite{Garg2011}: in this proof-of-concept, the authors show that in the limit of small periods, the thermal conductivity of the SL can exceed that of both the constituent materials, as a result of a dramatic reduction in the scattering of acoustic phonons by optical phonons, leading to very long phonon lifetimes.

In conclusion, we have investigated thermal transport in defected \mos \ monolayers through molecular dynamics simulations by realizing different distributions of vacancies, from random arrangement to repeated patterns of ordered strips of defects. While random distributions result in an overall suppression of thermal transport, we obtain the non-trivial result of realizing a SL which thermal conductivity matches the one of pristine \mos. When the super periodicity of these vacancies is achieved through $n_s=2$, regardless of system length and of the width of the defected region, $\kappa$ shows essentially the same behavior of pristine \mos. {By means of lattice dynamics calculations, we show that this SL periodicity endows the \mos \ monolayer with metamaterial-like properties. By modulating phonon band structure, it leads to a suppression of scattering via localized vibrational modes,  enabling essentially unperturbed acoustic phonon modes and group velocities}. 

\begin{acknowledgement}
We acknowledge the financial support under the National Recovery and Resilience Plan (NRRP), Mission 4 Component 2 Investment 1.3 - Call for tender No.341 published on March 13, 2022) by the  Italian Ministry of University and Research (MUR) funded by the European Union – NextGenerationEU. Award Number: Project code PE00000021, Concession Decree No. 1561 adopted on October 11, 2022 by the Italian Ministry of Ministry of University and Research (MUR), CUP F53C22000770007, Project title "NEST - Network 4 Energy Sustainable Transition". F.S. acknowledges that this publication was produced while attending the PhD programme in Physics at the University of Cagliari, Cycle XXXVIII, with the support of a scholarship financed by the Ministerial Decree no. 351 of 9th April 2022, based on the NRRP - funded by the European Union - NextGenerationEU - Mission 4 "Education and Research", Component 1 "Enhancement of the offer of educational services: from nurseries to universities" - Investment 4.1 "Extension of the number of research doctorates and innovative doctorates for public administration and cultural heritage".
\end{acknowledgement}

\section*{Data Availability Statement}
The data that support the ﬁndings of this study are available from the corresponding author upon reasonable request.

\bibliography{MoS2}

\end{document}